\documentclass[reprint,eqsecnum,floats,aps,amsmath,amssymb,nofootinbib,onecolumn,doublespacing,showpacs]{revtex4-2}

\usepackage{graphicx}
\usepackage{amsmath,amssymb}
\usepackage{hyperref}
\usepackage{graphicx}
\usepackage{subfigure}
\usepackage{arydshln}
\usepackage{inputenc}
\usepackage{graphicx}
\usepackage{amsmath,amssymb}
\usepackage{hyperref}
\usepackage{appendix}
\usepackage{color}

\begin{document}

\title{Analytic Primordial Power Spectrum in the Dressed Metric Approach to Loop Quantum Cosmology and Thermodynamics of Spacetime}

\author{Ana Alonso-Serrano,}
\email{ana.alonso.serrano@aei.mpg.de}
\affiliation{Institut für Physik, Humboldt-Universität zu Berlin, Zum Großen Windkanal 6, D-12489 Berlin, Germany\\
Max-Planck-Institut f\"ur Gravitationsphysik (Albert-Einstein-Institut), Am M\"{u}hlenberg 1, 14476 Potsdam, Germany}
\author{Guillermo A. Mena Marug\'an,}
\email{mena@iem.cfmac.csic.es}
\affiliation{Instituto de Estructura de la Materia, IEM-CSIC, Serrano 121, 28006 Madrid, Spain}
\author{Antonio Vicente-Becerril}
\email{antonio.vicente@iem.cfmac.csic.es}
\affiliation{Instituto de Estructura de la Materia, IEM-CSIC, Serrano 121, 28006 Madrid, Spain}

\begin{abstract}
We investigate the primordial power spectrum of cosmological tensor perturbations in the dressed metric approach to Loop Quantum Cosmology. We compute the background-dependent effective mass that affects their propagation using the effective description of Loop Quantum Cosmology and show that this mass can be approximated in different cosmological epochs by appropriate analytic functions. Moreover, in each of those epochs we can analytically solve the propagation of the perturbations, then obtaining the general solution globally by continuity requirements. On the other hand, since there are regimes far away from slow roll in the considered background evolution, the Bunch-Davies state does not provide a privileged choice of vacuum that would pick out a specific solution for the perturbations. Instead, we select the state of these perturbations by a recently proposed criterion that removes unwanted oscillations in the power spectrum. We compute the spectrum of this vacuum and compare it with other spectra obtained in the literature, especially with one corresponding to the hybrid approach to Loop Quantum Cosmology. Finally, we notice that the same type of background dynamics is found in a phenomenological approach to quantum gravity based on thermodynamics, allowing in this case a free value for the tantamount of the critical density. Extending the dressed metric proposal to this phenomenological model, one might expect a similar form for the associated primordial power spectrum.
\end{abstract}


\maketitle

\section{Introduction}

One of the major challenges of modern physics is to construct a consistent and complete quantization of General Relativity (GR). Among the most solid candidates to obtain such a quantum theory, Loop Quantum Gravity (LQG) is one of the most prominent proposals. It is a nonperturbative quantization program that was originally based on a canonical formulation of Einstein gravity in terms of the Ashtekar variables \cite{LQG,Thie}. These variables are a gauge connection and a densitized triad, that capture the physical information about the spatial geometry and its extrinsic curvature. Some of the most remarkable achievements of LQG have been attained by its application to cosmology, founding a discipline that is now called Loop Quantum Cosmology (LQC) \cite{LQC}. For instance, it has been shown that LQC solves the cosmic singularity problem, replacing the {\it{Big Bang}} with a cosmic bounce, often referred to as the {\it{Big Bounce}} \cite{LQC,APS}. This bounce provides a quantum transition from a contracting phase of the Universe to an expanding one.

To falsify any theory of quantum gravity, one should find a way to confront the predictions of that theory with observations. Among the most important observables at our disposal in cosmology, we find the cosmic microwave background (CMB) \cite{Planck}. The distribution of temperatures and polarization anisotropies in the CMB can be well explained by the existence of quantum primordial perturbations, which were originated in a vacuum state and experienced the effects of an inflationary epoch \cite{Mukhanov1}. If slow-roll inflation is a good approximation to describe that epoch, so that the background dynamics is quasi-de Sitter, then there is a preferred vacuum state for the perturbations, namely the Bunch-Davies state \cite{Bunch}. This state can be characterized as the unique Hadamard state that is invariant under the de Sitter isometry group \cite{Mukhanov1,Bunch}. Therefore, invariance and a good behavior of the correlation functions select a natural initial state in this case. 

However, this scenario does not account for the existence of a pre-inflationary period with radically different dynamics, such as the presence of a cosmological singularity in the context of GR or a bounce in LQC. In these cases, the quantum cosmological perturbations may be influenced by the dynamics of these earlier periods. For the affected perturbation modes, the preference for a Bunch-Davies state as the natural vacuum becomes meaningless. If those modes are observable today (e.g. because inflation was short-lived and they were not swept away, far beyond the present Hubble horizon), we need some alternative criterion to choose a suitable vacuum for them \cite{NBM}.

Selecting a vacuum is akin to fixing initial conditions. Different criteria to pick them out then lead to distinct solutions for the perturbations and, consequently, to different power spectra. To address this issue, several criteria have been proposed in the literature in order to identify vacuum candidates. The simplest prescription is the use of adiabatic states \cite{Parker}. Other proposals have explored the minimization of quantities related to the renormalized stress-energy tensor \cite{Agullo1,Lueders,Handley}, a minimal quantum uncertainty around the bounce compatible with classical properties at the end of inflation \cite{AG1,AG2}, a minimization of power oscillations during a period that covers a kinetically dominated epoch (where the inflaton potential has negligible contribution) and the beginning of inflation \cite{deBlas}, or an asymptotic diagonalization of the Hamiltonian of the perturbations in the ultraviolet sector \cite{NMP,NMT,NMP2}. In this work, we adopt this last approach to determine our vacuum state. 

It has been shown that the vacuum selected by this diagonalization also leads to spectra with non-oscillating properties \cite{NMP}. For this reason, it is often called the non-oscillating vacuum state. Nonetheless, this nomenclature can create some confusion with the original non-oscillating vacuum suggested by Mart\'{\i}n-de Blas and Olmedo \cite{deBlas}, which shows properties that depend crucially on the kinetically dominated period where the background has classical dynamics and the quantum effects are ignorable. This much more classical vacuum has been studied in Refs. \cite{Benito,Benito2}, and bears a resemblance to the vacuum put forward in Ref. \cite{Contaldi} within pure GR. To make clear the distinction between these different vacua, and emphasize in this way that we are interested in studying states that are well adapted to quantum effects, we will refer to the non-oscillating (NO) vacuum determined by asymptotic Hamiltonian diagonalization (AHD) \cite{NMT} as the NO-AHD vacuum.

The aim of this paper is to compute the primordial power spectrum (PPS) of cosmological perturbations using the dressed metric formalism. According to this formalism, the equations of motion of the perturbations in GR are modified by dressing the background cosmology with corrections coming from quantum geometry phenomena \cite{dressed1,dressed2,dressed3}. These corrections can be encapsulated in a number of background expectation values of geometric quantities, that can be evaluated on effective trajectories for sufficiently peaked quantum states. Within this approach, we will show how to obtain the solutions of the perturbations at the end of inflation analytically thanks to a number of well-motivated approximations in the description of the background evolution. The computation of the PPS then will become almost direct. This spectrum can be compared with others found in the literature, especially with the spectrum obtained in GR and with the spectrum associated with an alternative prescription for the quantization of the perturbations (using the same criteria for the choice of vacuum state \cite{NM}) in the framework of LQG, called hybrid LQC \cite{hybrid,MMO,CMM,CMBO,Bao}. This last prescription combines a loop quantization of the background geometry with a Fock quantization of the perturbations, treating the union of these two ingredients as a  constrained gravitational system \cite{hybrid}. Our comparison will give us a new perspective about the impact that different quantization procedures have on the power spectrum. 

Remarkably, our study can be extended to a phenomenological approach to quantum gravity based on the thermodynamics of spacetime \cite{Alonso3}.  In fact, this approach recovers the same cosmological solutions found in LQC  if a free phenomenological parameter of the resulting model is fixed to a suitable value. This coincide strongly motivates the idea of extending the quantization techniques used in LQC for cosmological perturbations, which are fully justified in the framework of LQG for the aforementioned value of the model parameter, to other values, at least in a reasonable neighborhood. This analysis can always be considered as a rigorous quantization of the perturbations in LQC around FLRW cosmological backgrounds if the critical density is permitted to vary. This variation can be achieved, e.g., by letting the Immirzi parameter free in LQG \cite{LQG}. As one could expect, the preferred value for the critical density in LQC \cite{LQC} (obtained with a value of the Immirzi parameter determined by black hole entropy arguments) is of Planck order. Other values lead to different bouncing cosmologies, that may also be related to other models with a quantum gravity basis (see e.g. Refs. \cite{Kaul,Meissner_2004}). This novel perspective allows us to apply the results of this work, particularly the form of the PPS, to this new phenomenological framework for which the implications remain largely unexplored.

The structure of this paper is as follows. In Sec. 2 we very briefly review the effective LQC background that we consider, we introduce cosmological perturbations paralleling the treatment in GR, and we present dynamical equations for the perturbation modes in this setting. The obstructions to solve analytically these dynamics of the perturbations lead us to suggest a division of their pre-inflationary and inflationary evolution into four epochs. We approximate the background-dependent mass of the perturbations in each of these epochs in such a way that we can derive the general solution for the perturbation modes. We then dedicate Sec. 3 to the definition of the NO-AHD vacuum and its application to determine the initial state of the perturbations. In Sec. 4, we calculate the power spectra, discussing the main results and the differences with respect to the hybrid prescription for the quantum dynamics of the perturbations in the framework of LQC. Finally, Sec. 5 contains the conclusions. In the Appendix, we review the key ingredients of the modified cosmology derived from thermodynamics of spacetime, pointing out its similarity with LQC. In the following, we use Planck units, setting $G$, $c$, and $\hbar$ (and also the Boltzmann constant) equal to one.

\section{Cosmological perturbations in the dressed metric prescription}

In this section, we will provide the background equations of a Friedmann-Lemaître-Robertson-Walker (FLRW) cosmology with a matter scalar field in the effective description of LQC. We will also introduce the propagation equations of the cosmological perturbations on the corresponding effective background, adopting the dressed metric prescription within LQC.

\subsection{Background in Effective Loop Quantum Cosmology}
\label{modified cosmology}

We consider homogeneous and isotropic FLRW spacetimes containing a free scalar field $\phi(t)$, corresponding to the kinetically dominated regimes that are found around the bounce in the phenomenologically interesting backgrounds of effective LQC. In this effective description and restricting our attention to such regimes, the dynamics of these background cosmologies are determined by the following modified Friedmann equation \cite{LQC}:

\begin{eqnarray}\label{eq::Modified_Friedmann_free_field}
H^2 = \frac{4\pi  }{3} \dot{\phi}^2  \Big [1 - 2\pi D  \dot{\phi}^2 \Big] ,
\end{eqnarray}
where  $D = 1/(4 \pi  \rho_\text{c} )$. Here, $\rho_c$ is the critical density. This density is an upper bound on the energy density of the inflaton, and it is reached precisely at the bounce that all effective trajectories display \cite{LQC,APS}. More concretely, $\rho_\text{c} =3/(8\pi  \gamma^2 \Delta)$, where $\gamma$ is the so-called Immirzi parameter \cite{Immirzi}, for which we take the usual value $\gamma \approx 0.2375$ motivated by black hole entropy arguments, and $\Delta= 4\sqrt{3}\pi \gamma $ is the area gap, which provides the minimum nonzero eigenvalue allowed by the spectrum of the area operator in LQG \cite{Thie}. The parameter $D$ can be considered a free real quantity if one allows the critical density to vary, e.g. if one decides not to fix the value of the Immirzi parameter. A similar situation is encountered in the framework of thermodynamics of spacetime, where one actually finds the same type of background dynamics with a free value of the parameter $D$. For more details, we refer to the Appendix.

On the other hand, taking into account local energy conservation, one can solve the Friedmann equation analytically, obtaining the following solution for the scale factor:
\begin{eqnarray}
a^6 = a_\text{B}^6 \Bigg[ 1+\frac{6}{D} \Big (t-t_\text{B} \Big)^2 \Bigg ],\label{eq::Fast_roll_Mod_a}
\end{eqnarray}    
where $a_\text{B} $ is the minimum of the scale factor and $t_\text{B}$ is the value of the proper time at the bounce. The Hubble parameter becomes
\begin{eqnarray}
H(t) = \frac{t-t_\text{B}}{3(t-t_\text{B})^2 + D /2} .\label{eq::Mod_H_Param}    
\end{eqnarray}
As we have commented, the appearance of a minimum $a_B$ indicates the occurrence of a bounce. There, a contracting phase of the Universe ($H < 0$) joins an expanding one ($H > 0$), which is the branch of observational interest. Note also that the value of $a_B$ can always be set equal to one by rescaling the scale factor in our flat FLRW cosmology.

\subsection{Cosmological perturbations}

In the standard cosmological model, the Universe experiences an inflationary period, which for simplicity we assume that is driven by a scalar field, that we can view as the inflaton. This accelerated expansion amplifies the quantum fluctuations in vacuo, so that they become macroscopic cosmological fluctuations \cite{Langlois1}. We consider a simplified scenario where the background dependence of the perturbation equations in GR remains valid. Therefore, the effective background-dependent mass that affects the perturbations has the same form as in Einstein gravity. However, instead of using the FLRW background of GR, we use the modified background presented in the previous section. In the context of LQC, a scenario of this kind has been proposed to describe the modified dynamics of the perturbations owing to the quantum corrections that effectively dress the background geometry. In general, the quantum effects on the background are encapsulated in a number of expectation values, that for highly peaked states can be obtained by evaluation on the effective trajectory of the peak. This proposal is known in LQC as the dressed metric prescription \cite{dressed1,dressed2,dressed3}. Adopting it, the Fourier modes $u_{\vec{k}}$ of the tensor perturbations turn out to satisfy an equation of the form \cite{NBMmass}
\begin{eqnarray}\label{udynamics}
u_{\Vec{k}}'' + \left(k^2 + s^{(t)}\right)u_{\Vec{k}} = 0,
\end{eqnarray}
where $\vec{k}$ is the wavevector, the wavenumber $k$ is its norm,  and $s^{(t)}$ plays the role of a background-dependent mass. For scalar perturbations, the analog equation is known as the Mukhanov-Sasaki equation, and the corresponding background-dependent mass differs in additional contributions arising from the inflaton potential \cite{Mukhanov2,Sasaki1,Sasaki2}. In our case, nonetheless, we will ignore those contributions for simplicity, describing the inflaton as if it were a massless field. This is a helpful abstraction that can be maintained in all epochs where the energy density of the inflaton is dominated by its kinetic part. This simplification leads in practice to the consideration of dynamical equations and PPS that correspond to tensor perturbations \cite{NBMmass}. 

To confront predictions of different models with observations, the CMB is one of the better cosmological observables at our disposal. In this task, the two-point functions play a key role (for instance, by measuring the temperature anisotropies in the sky we obtain the two-point function of the temperature distribution, which can be related to the angular distribution of the scalar perturbations). In turn, the two-point functions are determined by the Fourier transform of the PPS divided by the cube of $k$ (up to a constant global factor). In the tensor case that we are considering, the PPS is given by 
\begin{eqnarray} 
\mathcal{P}_\text{T}(k) = \frac{32 k^3}{\pi} \frac{|u_{\vec{k}}|^2}{a^2}. 
\end{eqnarray}
This PPS depends on the background scale factor and on $k$, rather than on $\vec{k}$, because Eq. \eqref{udynamics} implies that the dynamics of the modes $u_{\vec{k}}$ depends just on their wavenumber, but not on the direction of the wavevector. We will make this fact manifest by changing our notation for the modes from $u_{\vec{k}}$ to $u_k$ in the following. Besides, all quantities in the above formula must be evaluated at the end of inflation, where the PPS is defined. To compute this spectrum, we then need to integrate the dynamical equations of the perturbations starting with their initial data, and determine in this way the mode solution $u_{k}$.\\

\subsection{Effective mass for the dressed metric prescription}

In this subsection, we will compute the effective mass of the perturbations. At leading order in perturbation theory, the equations of motion for our gauge-invariant perturbations in GR take the form \eqref{udynamics} with a background-dependent mass given by $ s^{(t)}=-a''/a$ \cite{Langlois1,Langlois2}. We assume the validity of these equations even if the scale factor $a$ does not follow the dynamical evolution corresponding to GR, but instead evolves according to the effective dynamics of LQC. To obtain the explicit expression of the equations of the perturbations, we have to compute the effective mass in conformal time, using the scale factor \eqref{eq::Fast_roll_Mod_a} that we found by solving the modified Friedmann equation. In this way, we obtain
\begin{eqnarray}\label{eq::Effective_Mass_exact}
s^{(t)}= \frac{2a_B^2}{D^{1/3}} \frac{2(t-t_\text{B})^2-D}{[6(t-t_\text{B})^2+D]^{5/3}}.
\end{eqnarray}
However, the above formula is given in proper time $t$, while what we need to know is the effective mass in terms of the conformal time. 

The relationship between the cosmic and the conformal times (for $D>0$) is
\begin{eqnarray}
\eta - \eta_\text{B} = \int^t_{t_\text{B}} \frac{dt}{a(t)} = \frac{(t-t_\text{B})}{a_\text{B}} \  _2F_1\Big(\frac{1}{6},\frac{1}{2};\frac{3}{2};-\frac{6(t-t_\text{B})^2 }{D}\Big). 
\end{eqnarray}
The appearance of the hypergeometric function $_2F_1$ does not allow for an explicit inversion, preventing the derivation of an analytically tractable expression for the effective mass. This makes impracticable the resolution of Eq. \eqref{udynamics} for the perturbations. To circumvent this obstruction, we will propose below an approximation for the considered, time-dependent effective mass. In addition, we will incorporate the effects of inflation in a simplified manner by including a period of de Sitter expansion. In total, we will divide the evolution of the perturbation modes into four epochs, each of them characterized by a different background behavior and its corresponding effective mass. To connect these different epochs, we will demand continuity up to the first derivative on the background scale factor, as well as on all of the perturbation modes. These requirements will fix the matching of the solutions at the end times of the consecutive epochs.  

The first epoch corresponds to the period with important modifications with respect to GR that result in a bounce of the background geometry. In this period, the effective mass can be approximated by a P\"oschl-Teller potential \cite{NM,PT1,PT2,PT3,Bao2} plus the addition of a constant term. After this initial interval, we will consider a kinetically dominated epoch in which the background can be accurately described using GR, with matter given by a massless scalar field \cite{Contaldi,NBM}. In effective LQC, this kinetic regime appears in all the physically interesting background solutions, for which the subsequent quantum imprints on the perturbations are compatible with the observations but not totally negligible \cite{Morris}. To achieve the best possible approximation, we will introduce a transition interval where the effective mass is constant. This interval will be followed by a genuine period of kinetic dominance. As the scalar field gradually loses its kinetic energy, it starts to feel the inflaton potential, which finally becomes dominant, making way to an inflationary phase. To simplify the description of this last epoch, we will adopt a de Sitter expansion, as we commented above. The analytic expression of the effective mass in each of these epochs is as follows.\\

\begin{figure}
\centering
\includegraphics[width=18cm]{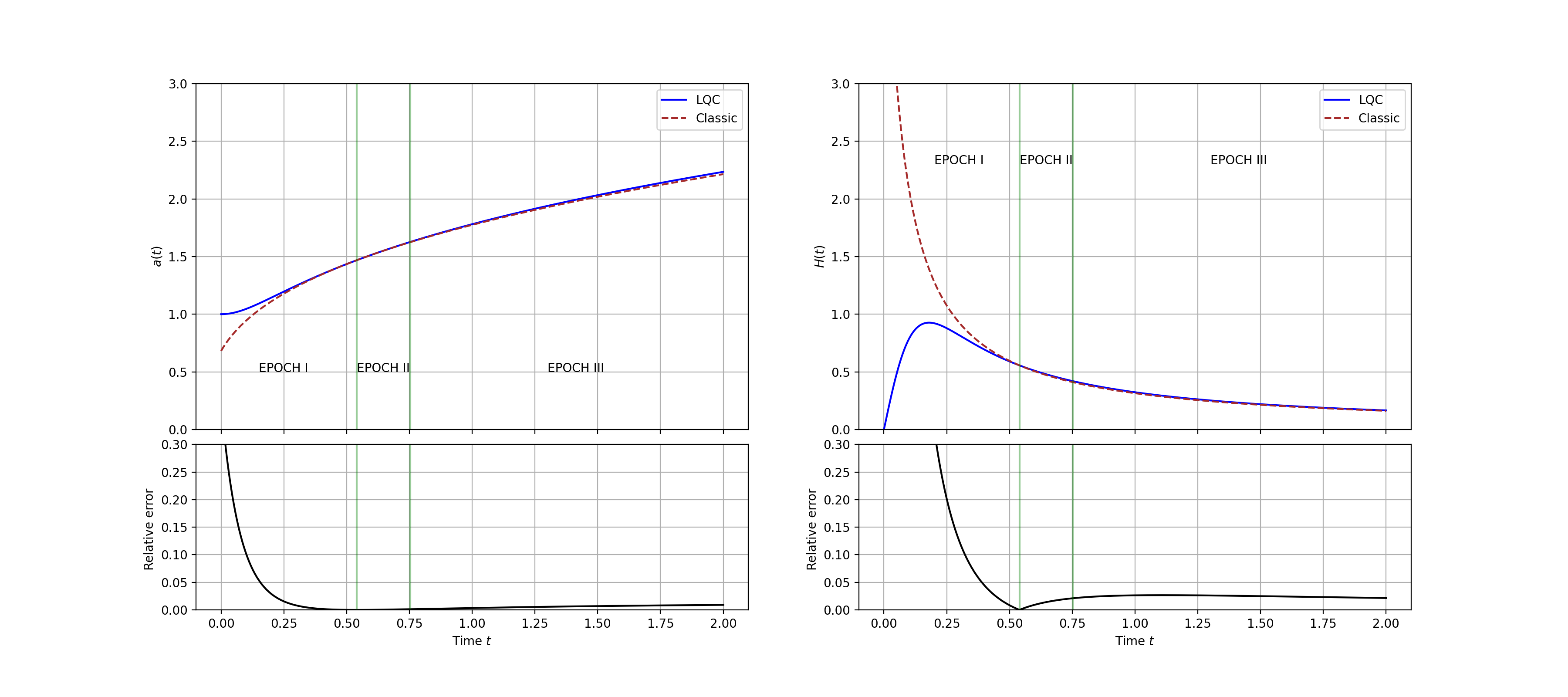}
\caption{Top left: evolution of the scale factor in the dressed metric prescription for LQC (blue) and in GR for kinetic dominance (red). We have set the value of the scale factor at the bounce equal to one. Top right: evolution of the Hubble parameter in the dressed metric prescription for LQC (blue) and in GR for kinetic dominance (red). The green vertical lines separate the different epochs. At the bottom, we display the respective relative errors (in absolute value) with respect to the average of the two compared quantities. } 
\label{fig::a_scalar_wth_error}
\end{figure}

\begin{itemize}
\item{\textbf{First approximation: P\"oschl-Teller potential.}}
This first epoch covers an interval $[t_\text{B}, t_0 ]$, where $t_\text{B}$ is the proper (or cosmic) time at the bounce. The P\"oschl-Teller potential with an added constant that is used to approximate the effective mass in this interval has the form 
\begin{equation}\label{eq::Effective_mass_PT}
s_{\text{PT}}^{(t)}(\eta_0)= \frac{U_0-v_0}{\cosh^2\Big({\alpha(\eta-\eta_\text{B})}\Big)} + v_0.
\end{equation}
Here, $U_0$, $v_0$, and $\alpha$ are just constants. To determine them, we use the exact effective mass given by Eq. \eqref{eq::Effective_Mass_exact}. We choose three specific moments: The bounce ($t=t_\text{B}$), the instant where the exact effective mass becomes null ($t_\text{null}=\sqrt{D/2} + t_\text{B}$), and the instant where the exact effective mass has a maximum, that we choose as the endpoint of the interval ($t_{0}=\sqrt{3D/2}+t_\text{B}$). We then require that the P\"oschl-Teller and the effective masses coincide at these three values of the time $t$.

From the condition at the bounce, we obtain the value $U_0 = -2/D$. By considering the other two conditions, we derive two equations that we solve numerically to fix the values of $v_0$ and $\alpha$. The background evolves using Eqs. \eqref{eq::Fast_roll_Mod_a} and \eqref{eq::Mod_H_Param}.
   
\item{\textbf{Second approximation: constant effective mass.}} 
This epoch covers an interval $(t_0, t_0 ' )$, with an end point $t_0'$ that will be specified below (see Eq. \eqref{t0prime}). This epoch has a constant effective mass for the perturbations. This constant mass equals the value of the P\"oschl-Teller approximation at the matching time $t_{0}$. Thus
\begin{eqnarray} \label{eq::Effective_mass_cste}
s_{0}^{(t)} (\eta) =  \frac{U_0-v_0}{\cosh^2\Big({\alpha(\eta_{0}-\eta_\text{B})}\Big)} + v_0 = m_0,
\end{eqnarray}
where $\eta_0$ is the conformal time corresponding to $t_0$. The evolution of the background in this period is governed by the Friedmann equation corresponding to kinetic dominance (in Fig. \ref{fig::a_scalar_wth_error}, we can see that the relative error in the background between the LQC cosmology and the classical cosmology in GR becomes negligible). Therefore, the scale factor adopts the form 
\begin{eqnarray}
a_\text{class}(t)=a_\text{R} \left( \sqrt{12 \pi } (t-t_\text{R}) \right)^{1/3}.
\end{eqnarray}
Here, $a_\text{R}$ and $t_\text{R}$ are two constants chosen to ensure continuity up to the first derivative at the matching time $t=t_0$, where we have in particular $a(t_0)=a_0$. In this way, we get 
\begin{eqnarray}\label{eq::a_scale_kinetic}
a_0= a_\text{R} \left(\sqrt{12\pi  } (t_0-t_\text{R})\right)^{1/3}=a_\text{B} \left[1+ \frac{6}{D}\left(t_0-t_\text{B}\right)^2 \right]^{1/6},
\end{eqnarray}
and $t_\text{R} = t_\text{B} - D  /[6(t_0 - t_\text{B})]$.

\item{\textbf{Third approximation: kinetic dominance in GR.}} 
This epoch covers an interval $[t_0', t_i ]$. The end time $t_i$ will be estimated later on, and will be identified with the onset of inflation. In this period, the background and the effective mass can be taken as those corresponding to kinetic dominance in GR. Thus, the effective mass is  
\begin{eqnarray}\label{eq::scalar_effective_mass_class}
s_\text{class}^{(t)} &=& \frac{4\pi  a_\text{R}^2}{3} \left ( \sqrt{12 \pi } (t-t_\text{R}) \right )^{-4/3}.
\end{eqnarray}

To obtain the effective mass in conformal time, we first compute the relationship between the proper and conformal times
\begin{eqnarray}
\eta - \eta_0  &=& \frac{1}{2a_\text{R} H_0} \left[\left(\sqrt{12\pi } (t-t_{R})\right)^{2/3} - \left(\frac{a_0}{a_\text{R}}\right)^2 \right].\label{proper_to_conofrmal}
\end{eqnarray}
We have called $H_0=\sqrt{4\pi /3 }$. This relationship leads to
\begin{eqnarray}
s_\text{class}^{(t)} (\eta) = \frac{1}{4} \left[(\eta-\eta_0) + \left(\frac{a_0^2}{2a_\text{R}^3 H_0}\right) \right]^{-2} . \label{eq::Effective_mass_GR}
\end{eqnarray}

This epoch starts when the effective mass of this kinetically dominated regime becomes equal to the value of the constant effective mass of the preceding epoch, at a time $\eta_0'$ that is therefore given by the identity
\begin{eqnarray}\label{t0prime}
s_\text{class}^{(t)} (\eta_0')= \frac{1}{4} \left[(\eta_0'-\eta_0) + \left(\frac{a_0^2}{2a_\text{R}^3 H_0}\right) \right]^{-2} = \frac{U_0-v_0}{\cosh^2\left({\alpha(\eta_{0}-\eta_\text{B})}\right)} + v_0. 
\end{eqnarray}\\
Using relationship \eqref{proper_to_conofrmal} between the proper and conformal times, we can equivalently determine the corresponding value of $t_0'$ from the above equality.

\item{\textbf{Fourth approximation: de Sitter inflation.}} 
This last epoch covers an interval $(t_i, \infty )$. Its end point has been taken at infinity, but we can equally take any finite time where inflation has managed to freeze all the considered perturbation modes. In the literature of LQC, the conformal time $\eta_i$ is commonly set equal to $\eta_i \approx 1600$ for phenomenological reasons, since this is the typical value found in background solutions that allow for sufficient inflation while maintaining observable quantum modifications in the PPS \cite{Agullo2,NBM,NM}. After the kinetically dominated epoch, an inflationary period begins. For the sake of simplicity, we will consider that the transition between these two epochs is instantaneous, and we will describe the inflationary period as a de Sitter expansion, driven by a constant inflaton potential that plays the role of a cosmological constant \cite{Mukhanov1,Langlois1}. This constant can be identified (modulo a global factor) with the square of the Hubble parameter at the time $\eta_i$ when the inflationary epoch starts. The corresponding scale factor is 
\begin{eqnarray}
a_\text{dS}(\eta) = \left[a_i^{-1} - H_\Lambda (\eta-\eta_i)\right]^{-1},     
\end{eqnarray}
where $H_\Lambda=H_\text{clas}(\eta_i)$ and $a_i=a_\text{clas}(\eta_i)$, so that the scale factor is continuous up to its first derivative. The effective mass of the perturbations becomes
\begin{eqnarray}\label{eq::Effective_mass_dS}
s^{(t)}_\text{dS}(\eta) = -2H_\Lambda^2 \left[a_i^{-1}-H_\Lambda(\eta-\eta_i)\right] ^{-2}.
\end{eqnarray}
\end{itemize}

\begin{figure}[h!]
\centering
\includegraphics[width=16cm]{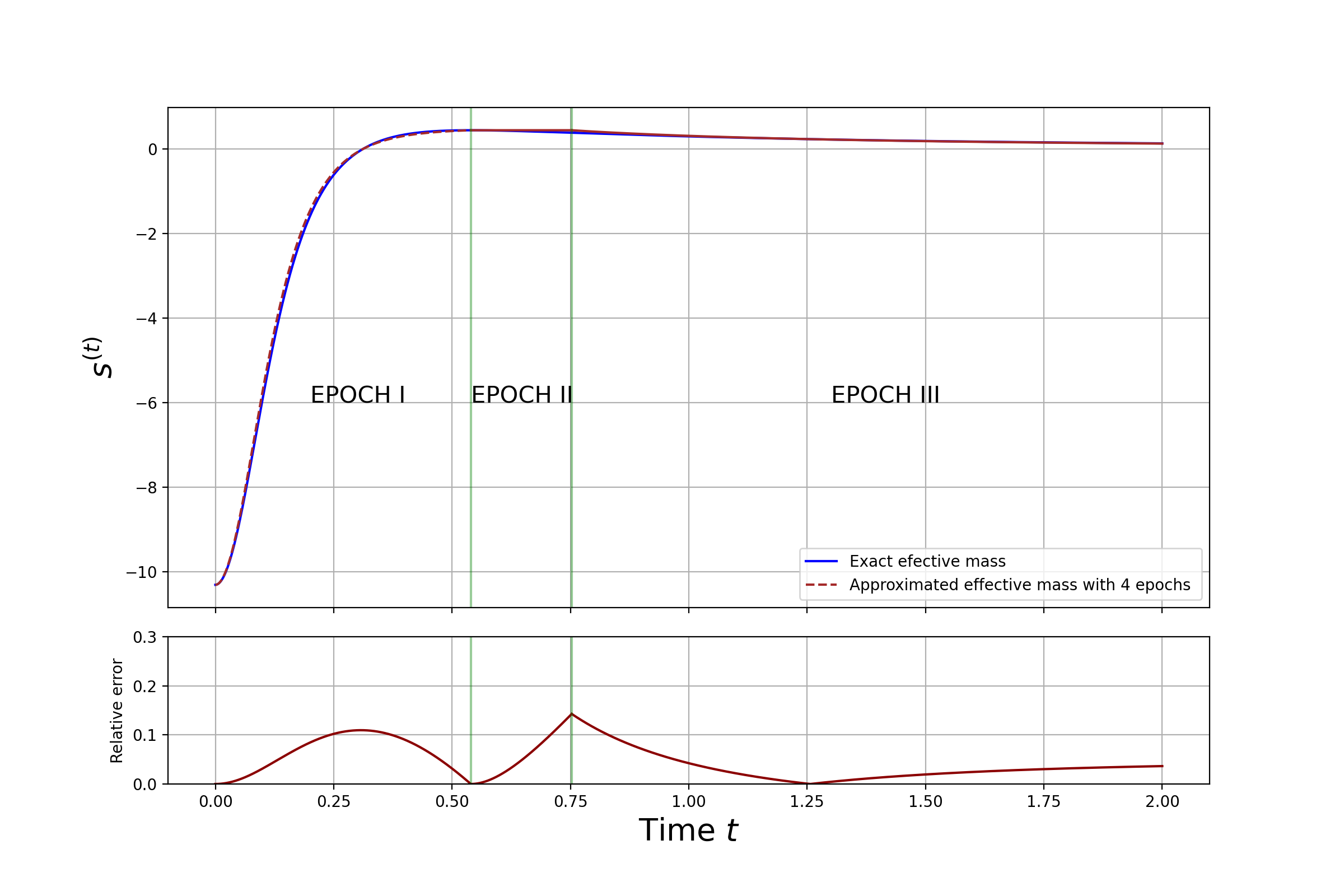}
\caption{Top: evolution of the exact effective mass (blue) and of the approximated effective mass (red) with 4 different epochs. The green vertical lines separate these epochs. Bottom: relative error in the effective masses.}
\label{fig::s_with_error_4zones}
\end{figure}

In order to check the consistency of our approach, we display in Fig. \ref{fig::s_with_error_4zones} the effective mass for the first three epochs compared with the exact, background-dependent effective mass given by Eq. \eqref{eq::Effective_Mass_exact}, as well as the relative error between them. It is clear that the approximation works very well (see e.g. the discussion in Ref. \cite{NM}), with a relative error that is consistently below $15\%$. 

\subsection{Solution to the mode equation} 
In this subsection, we will solve the equations of motion of the perturbation modes in the different epochs of the evolution.

\subsubsection{P\"oschl-Teller}

The dynamical equation for the modes in the P\"oschl-Teller epoch can be written using Eq. \eqref{eq::Effective_mass_PT} as
\begin{equation}\label{eq::MS_bounce}
u_{k}''+ \left[\Tilde{k}^2+\frac{\alpha^2\lambda(\lambda-1)}{\cosh^2\alpha (\eta-\eta_\text{B})}\right]u_{k} =0 ,
\end{equation}
where $\lambda$ is defined by $U_0-v_0=\alpha^2\lambda(\lambda-1)$ and $\Tilde{k}^2=k^2 + v_0$. We introduce a new variable $x$ and a mode function $\nu_{\Tilde{k}}(x)$ given by \cite{PTCevik}
\begin{eqnarray}
x(\eta)&=&\frac{\tanh\alpha (\eta-\eta_\text{B}) +1}{2}=\left[1+e^{-2\alpha(\eta-\eta_\text{B})}\right]^{-1},\\ \nu_{\Tilde{k}}(x)&=&x^{-r_{\Tilde{k}}}(1-x)^{-s_{\Tilde{k}}}u_{k}(x)\,,
\end{eqnarray}
where $r_{\Tilde{k}}$ and $s_{\Tilde{k}}$ are 
\begin{equation}
r_{\Tilde{k}}=\frac{i \Tilde{k}}{2\alpha}, \qquad s_{\Tilde{k}} =-\frac{i\Tilde{k}}{2\alpha} .
\end{equation}
Therefore, Eq. \eqref{eq::MS_bounce} becomes 
\begin{equation}
x(1-x)\,\frac{d^2\nu_{\Tilde{k}}}{dx^2}+\left[\frac{i\Tilde{k}}{\alpha}-2x+1\right]\frac{d\nu_{\Tilde{k}}}{dx}+\lambda(\lambda-1)\nu_{\Tilde{k}}=0.
\end{equation}
Comparing this with the general form of a hypergeometric equation, 
\begin{eqnarray}\label{hyperg}
x(1-x)\ \frac{d^2\nu_{\Tilde{k}}}{dx^2}+\left[c-(a+b+1)x\right]\frac{d\nu_{\Tilde{k}}}{dx}-ab\nu_{\Tilde{k}}=0,
\end{eqnarray}
we see that, in our case, $a=\lambda$, $b=1-\lambda$, and $c=i(\Tilde{k}/\alpha)+1$. 

The general solution to Eq. \eqref{hyperg} is a linear combination of two hypergeometric functions, $_2F_1(a,b;c;x)$ and $x^{1-c}\,_2F_1(a-c+1,b-c+1;2-c;x)$ (provided that $c$ is not an integer \cite{Abra}). Therefore, the general solution of our dynamical equation \eqref{eq::MS_bounce} is 
\begin{eqnarray}
\notag u_{k} &=& A_k \left(\frac{x}{1-x}\right)^{i\Tilde{k}/2\alpha}{}_{2}F_{1}\left(\lambda,1-\lambda;\frac{i\Tilde{k}}{\alpha}+1;x\right) \nonumber \\
&+&  B_k \left[x(1-x) \right]^{-i\Tilde{k}/2\alpha}  {}_{2}F_{1}\left(\lambda-\frac{i\Tilde{k}}{\alpha},1-\lambda-\frac{i\Tilde{k}}{\alpha};1-\frac{i\Tilde{k}}{\alpha};x\right),
\end{eqnarray}
where $A_k$ and $B_k$ are arbitrary constants.

\subsubsection{Constant effective mass}

Let us consider now the epoch when the background dynamics are in a kinetically dominated regime and the effective mass of the perturbations can be approximated by a constant. From Eq. \eqref{eq::Effective_mass_cste}, the dynamical equations of the perturbation modes adopt the form 
\begin{eqnarray}
u''_k + \left( k^2 + m_{0}\right) u_k=0,
\end{eqnarray}
where $m_{0}$ is the constant value of the effective mass. The general solution to this mode equation can be written as
\begin{eqnarray}
u_k(\eta) = \alpha_k e^{i\kappa\eta  } + \beta_k  e^{-i\kappa\eta},
\end{eqnarray}
where $\kappa=\sqrt{k^2 + m_{0}}$.

To fix the constants $\alpha_k$ and $\beta_k$, we use the continuity up to the first derivative of the perturbation in the matching time with the P\"oschl-Teller epoch. In this way, we obtain
\begin{eqnarray}
\alpha_k&=&   \frac{e^{-i\kappa\eta_0}}{2\kappa}\left[\kappa u_k(\eta_0) -i u_k'(\eta_0)\right] ,\label{eq::alpha_k_cste}  \\
\beta_k&=& \frac{ e^{i\kappa\eta_0}}{2\kappa}\left[\kappa u_k(\eta_0)+i u_k'(\eta_0)\right]\label{eq::beta_k_cste}.
\end{eqnarray}

\subsubsection{Kinetic dominance}

In this epoch, the background continues in a kinetically dominated regime, while the effective mass of the perturbation changes according to the standard dependence on the background found in GR  \eqref{eq::Effective_mass_GR}. The mode equations become
\begin{eqnarray}
u_{k}''+ \left[k^2+  \frac{1}{4} \left( (\eta-\eta_0) + \left(\frac{a_0^2}{2 a_\text{R}^3 H_0}\right) \right)^{-2}  \right]u_{k} =0.\label{Mukhanov-Sasaki_GR}
\end{eqnarray}
We recall that $H_0= \sqrt{4\pi /3}$. Let us introduce the variables $\bar{y}= \eta-\eta_0 +a_0^2/(2a_\text{R}^3H_0)$ and $\bar{x}=k\bar{y}$, together with the function $v_k=\bar{y}^{-1/2} u_k $. We then get
\begin{eqnarray}
\frac{d^2v_k}{d\bar{x}^2} + \frac{1}{\bar{x}} \frac{dv_k}{d\bar{x}} + v_k=0.
\end{eqnarray}
This corresponds to a Bessel differential equation of order $\nu=0$. Therefore, the general solution to the mode equation can be expressed in terms of Hankel functions in the compact form
\begin{eqnarray}
u_k (\bar{y}) = \sqrt{\frac{\pi \bar{y}}{4}} \Big[C_k H_0^{(1)} (k\bar{y}) + D_k H_0^{(2)}(k\bar{y}) \Big ].\label{eq::sol_pert_kin}
\end{eqnarray}

To fix the constants $C_k$ and $D_k$, we use again the continuity of the perturbations up to their first time derivative at the matching point with the previous epoch. These conditions lead to
\begin{eqnarray}
\notag C_k &=& \frac{ i }{4} \Bigg[\alpha_k e^{i\kappa\eta_0'} \left( \sqrt{\frac{\pi}{\bar{y}_0'}}\left(1 -2i\kappa \bar{y}_0'\right) H_0^{(2)}(k\bar{y}_0') - 2\sqrt{\pi \bar{y}_0' k^2} H_1^{(2)}(k\bar{y}_0')   \right) \\&+& \beta_k e^{-i\kappa\eta_0'} \left(  \sqrt{\frac{\pi}{\bar{y}_0'}}\left(1 +2i\kappa \bar{y}_0'\right) H_0^{(2)}(k\bar{y}_0') - 2\sqrt{\pi \bar{y}_0' k^2} H_1^{(2)}(k\bar{y}_0')   \right) \Bigg]\label{eq::C_cste},\\
\notag D_k &=&  -\frac{ i }{4}\Bigg[ \alpha_k e^{i\kappa\eta_0'}\left( \sqrt{\frac{\pi}{\bar{y}_0'}} \left(1 - 2i \kappa\bar{y}_0'\right) H_0^{(1)}(k\bar{y}_0') - 2\sqrt{\pi \bar{y}_0' k^2} H_1^{(1)}(k\bar{y}_0')  \right) \\&+&  \beta_k e^{-i\kappa\eta_0'} \left(   \sqrt{\frac{\pi}{\bar{y}_0'}} \left(1 +2 i \kappa\bar{y}_0'\right) H_0^{(1)}(k\bar{y}_0') - 2\sqrt{\pi \bar{y}_0' k^2} H_1^{(1)}(k\bar{y}_0') \right)\Bigg] ,   \label{eq::D_cste}
\end{eqnarray}
where $\bar{y}_0'=\eta_0' - \eta_0 +a_0^2/(2a_\text{R}^3H_0) $, $a_\text{R}$ is the constant introduced in Eq. \eqref{eq::a_scale_kinetic}, and we recall that $\kappa=\sqrt{k^2 + m_{0}}$.
    
\subsubsection{de Sitter inflation}

Finally, we will describe the inflationary epoch with a de Sitter background, while the effective mass of the perturbation modes is given by the corresponding mass in GR:
\begin{eqnarray}
s^{(t)}_{dS} = -2H_\Lambda^2 \Big[a_i^{-1}-H_\Lambda(\eta-\eta_i)\Big] ^{-2}.
\end{eqnarray}
We introduce again variables of the form $\breve{y}=\eta_i-\eta + 1/(a_iH_\Lambda)$ and $\breve{x}=k\breve{y}$, together with the function $w_k=\breve{y}^{-1/2}u_k$, to rewrite the mode equations as
a Bessel equation of order $3/2$
\begin{eqnarray}
\frac{d^2w_k}{d\breve{x}^2} + \frac{1}{\breve{x}} \frac{dw_k}{d\breve{x}} +\left[1-\frac{9}{4\breve{x}^2}\right] w_k=0.
\end{eqnarray}
Then, in terms of the conformal time, the general solution for the perturbation modes is \cite{Mukhanov1}
\begin{eqnarray}
\notag u_k &=& G_k \frac{e^{ik(\eta-\eta_i - a_i^{-1}H_\Lambda^{-1})}}{\sqrt{2k}} \left[1+\frac{i}{k(\eta-\eta_i- a_i^{-1}H_\Lambda^{-1})}\right]  \nonumber  \\
&+&F_k \frac{e^{-ik(\eta-\eta_i - a_i^{-1}H_\Lambda^{-1})}}{\sqrt{2k}} \left[1-\frac{i}{k(\eta-\eta_i- a_i^{-1}H_\Lambda^{-1})}\right]. \label{eq::MS_sol_dS}
\end{eqnarray}

To fix the constants $G_k$ and $F_k$, we use again the requirement of continuity up to the first time derivative with respect to the solution in the preceding epoch. In this way, we conclude that
\begin{eqnarray}
\notag F_k &=&  \frac{e^{-ik/a_iH_\Lambda}}{4}\sqrt{\frac{\pi k }{ a_i H_\Lambda}}  \Bigg[C_k \left[  H_0^{(1)}\left(\frac{k}{2a_iH_\Lambda}\right) -  H_1^{(1)}\left(\frac{k}{2a_iH_\Lambda}\right)\left(i+\frac{a_iH_\Lambda}{k}\right) \right] \\&+& D_k \left[H_0^{(2)}\left(\frac{k}{2a_iH_\Lambda}\right)-H_1^{(2)}\left(\frac{k}{2a_iH_\Lambda}\right)\left(i+\frac{a_iH_\Lambda}{k}\right) \right]\Bigg] ,\label{eq::F_k_deSitter} \\
\notag G_k &=&  \frac{e^{ik/a_iH_\Lambda}}{4} \sqrt{\frac{\pi k }{a_i H_\Lambda}} \Bigg[C_k \left[  H_0^{(1)}\left(\frac{k}{2a_iH_\Lambda}\right) +  H_1^{(1)}\left(\frac{k}{2a_iH_\Lambda}\right) \left(i-\frac{a_iH_\Lambda}{k}\right) \right] \\&+& D_k \left[H_0^{(2)}\left(\frac{k}{2a_iH_\Lambda}\right)+H_1^{(2)}\left(\frac{k}{2a_iH_\Lambda}\right)\left(i-\frac{a_iH_\Lambda}{k}\right) \right]\Bigg]
\label{eq::G_k_deSitter}.
\end{eqnarray}

Summarizing, using our matching conditions between consecutive epochs, we have shown that the solution at the end of inflation can be analytically determined provided that we have initial data to fix the constants $A_k$ and $B_k$ for each of the perturbation modes in the P\"oschl-Teller period.

\section{Initial conditions and vacuum state}

In quantum field theory in general curved spacetimes, there is no preferred vacuum. This is a serious problem because, away from very symmetric spacetimes, we need a proposal to choose a natural state for the perturbations. This issue becomes crucial in cosmology, since predictions depend strongly on this choice. Indeed, selecting a vacuum is equivalent to setting initial conditions. These initial conditions determine the evolution of the cosmological perturbations in the Very Early Universe. Here, we will adhere to the NO-AHD vacuum proposal \cite{NMT,NMP,NM2}. This proposal picks out a vacuum that is well adapted to the background dynamics and leads to a power spectrum with non-oscillating properties. Such vacuum can be determined by an asymptotic diagonalization of the Hamiltonian of the perturbations in the ultraviolet region of asymptotically large wavenumbers. This proposal has been applied successfully for Minkowski and de Sitter backgrounds, where it selects the Poincar\'e and the Bunch-Davies vacua, respectively \cite{NMT}. Moreover, it has also been applied to cosmological perturbations in kinetically dominated regimes in GR, as well as in hybrid LQC scenarios \cite{NM}.

\subsection{NO-AHD vacuum state}

The NO-AHD vacuum state is associated with a set of annihilation and creation operators, with an evolution that depends on the background and which is directly governed by an asymptotically diagonal Hamiltonian. In this diagonal Hamiltonian, all interacting terms between the perturbations have been systematically eliminated order by order in the ultraviolet regime as $k$ tends to infinity, as described in Ref. \cite{NMT}. 

The amplitude of the mode solutions corresponding to the NO-AHD vacuum of the perturbations exhibits small scale and time-dependent oscillations \cite{NMP,NM}. These mode solutions can always be expressed as \cite{NMT,NM}
\begin{equation}\label{muh}
u_k=\sqrt{-\frac{1}{2\text{Im}(h_k)}}e^{i\int_{\eta_0}^\eta d\tilde\eta \text{Im}(h_k)(\tilde{\eta})}	,
\end{equation}
where $h_k$ is a solution of the Riccati differential equation with background-dependent mass $s^{(t)}$:
\begin{equation} \label{eq::diff_prop}
h_{k}'=k^2+s^{(t)}+h_k^2.
\end{equation}

The NO-AHD vacuum state is characterized by an asymptotic expansion of the form
\begin{equation}\label{eq::asymph_prop}
kh_k^{-1}\sim i\left[1-\frac{1}{2k^2}\sum_{n=0}^{\infty}\left(\frac{-i}{2k}\right)^{n}\gamma_n \right],
\end{equation}
where $\gamma_n$ are $k$-independent functions of time that are fixed by the recurrence relations
\begin{equation}\label{eq::recursion_prop}
\gamma_{0}=s^{(t)},\qquad \gamma_{n+1}=-\gamma_{n}'+4s^{(t)} \left[\gamma_{n-1}+\sum_{m=0}^{n-3}\gamma_m \gamma_{n-(m+3)}\right]-\sum_{m=0}^{n-1}\gamma_m \gamma_{n-(m+1)}.
\end{equation} 

\subsection{NO-AHD vacuum at the bounce for the dressed metric prescription}

We will now employ the NO-AHD vacuum proposal to fix the exact values of the constants that specify the mode solution in the region around the bounce. In the interval $[\eta_\text{B},\eta_0]$ (corresponding to $[t_\text{B},t_0]$ in proper time) where we have applied a P\"oschl-Teller approximation for the effective mass, the differential equation \eqref{eq::diff_prop} becomes
\begin{eqnarray}
h_{\Tilde{k}}' = \Tilde{k}^2 + \frac{U_0}{\cosh^2{\alpha(\eta-\eta_B)}} +h_{\Tilde{k}}^2,
\end{eqnarray}
where $\Tilde{k}^2=k^2+v_0$. Recall that $v_0$ is the constant added to the genuine P\"oschl-Teller contribution. To solve this differential equation we introduce the variable $x=[1+e^{-2\alpha(\eta-\eta_B)}]^{-1}$ and the function \cite{NM}
\begin{eqnarray} \label{eq::var_chang_prop}
h_{\Tilde{k}}=i{\Tilde{k}}(1-2x)-2\alpha x(1-x) \frac{d}{dx} (\log{\upsilon_{\Tilde{k}}}) = i{\Tilde{k}}(1-2x)-\frac{2\alpha x(1-x)}{\upsilon_{\Tilde{k}}}\frac{d\upsilon_{\Tilde{k}}}{dx}.
\end{eqnarray}
In this way, we obtain the hypergeometric equation 
\begin{eqnarray}
x(x-1)\frac{d^2\upsilon_{\Tilde{k}}}{dx^2}+\left[b_3^{\Tilde{k}} - (b^{\Tilde{k}}_1 + b^{\Tilde{k}}_2 +1)x\right] \frac{d\upsilon_{\Tilde{k}}}{dx} - b_1^{\Tilde{k}} b_2^{\Tilde{k}} \upsilon_{\Tilde{k}} =0 ,
\end{eqnarray}
where the parameters $b^{\Tilde{k}}_i$ are 
\begin{eqnarray}
b_1^{\Tilde{k}}=\frac{1}{2} \left( 1+ \sqrt{1+\frac{4U_0}{\alpha^2}}\right) - \frac{i\Tilde{k}}{\alpha},\quad b_2^{\Tilde{k}}=b_1^{\Tilde{k}}- \sqrt{1+\frac{4U_0}{\alpha^2}}, \quad  b_3^{\Tilde{k}} = 1 - \frac{i\Tilde{k}}{\alpha}.
\end{eqnarray}
The general solution to the above hypergeometric equation is 
\begin{equation}\label{solhyper}
\upsilon_{\Tilde{k}}= M_{\Tilde{k}}\ {}_2F_1(b_1^{\Tilde{k}},b_2^{\Tilde{k}};b_3^{\Tilde{k}};x) + N_{\Tilde{k}}\  x^{i\Tilde{k}/\alpha}  \;{}_2F_1(b_1^{\Tilde{k}}- b_3^{\Tilde{k}}+1,b_2^{\Tilde{k}}-b_3^{\Tilde{k}}+1;2-b_3^{\Tilde{k}};x),
\end{equation}
where $M_{\Tilde{k}}$ and $N_{\Tilde{k}}$ are two constants.

We still need to impose on $h_{\Tilde{k}}$ the asymptotic behavior determined by Eqs. \eqref{eq::asymph_prop} and \eqref{eq::recursion_prop}. At dominant and first subdominant order in the asymptotic expansion, and using that $U_0 \cosh^{-2}{\alpha(\eta-\eta_b)}=4U_0 x(1-x)$, we get \cite{NM}
\begin{eqnarray}
h_{\Tilde{k}} =    -i \Tilde{k} \left[1 + \frac{2U_0}{\Tilde{k}^2} x(1-x) + \mathcal{O}(\Tilde{k}^{-3}) \right].
\end{eqnarray}
Furthermore, the terms $\mathcal{O}({\Tilde{k}}^{-3}) $ must depend polynomically on $x$ \cite{NM}. By incorporating this behavior into Eq. \eqref{eq::var_chang_prop} and integrating the result, we conclude that $\upsilon_{\Tilde{k}}$ admits an asymptotic expansion of the form 
\begin{eqnarray} \label{app_serie}
\upsilon_{\Tilde{k}}=x^{i{\Tilde{k}}/\alpha} \sum_{n=0}^\infty c_n^{\Tilde{k}} x^n ,
\end{eqnarray}
where $c_n^{\Tilde{k}}$ are complex constants that depend on linear combinations of inverse powers of ${\Tilde{k}}$. The only solution among those in Eq. \eqref{solhyper} that has an $x$-dependence of the form \eqref{app_serie} for all ${\Tilde{k}}$ is
\begin{eqnarray}
\upsilon_{\Tilde{k}}= \upsilon_0^{\Tilde{k}} x^{i{\Tilde{k}}/\alpha}  {}_2F_1(b_1^{\Tilde{k}}- b_3^{\Tilde{k}}+1,b_2^{\Tilde{k}}-b_3^{\Tilde{k}}+1;2-b_3^{\Tilde{k}};x),
\end{eqnarray}
with $\upsilon_0^{\Tilde{k}}$ a constant. Introducing it in Eq. \eqref{eq::var_chang_prop}, we obtain
\begin{eqnarray}
\notag h_{\Tilde{k}}&=&- i\Tilde{k} -2\alpha x(1-x) \frac{(b_1^{\Tilde{k}}-b_3^{\Tilde{k}}+1)(b_2^{\Tilde{k}}-b_3^{\Tilde{k}}+1)}{2-b_3^{\Tilde{k}}}  \;\nonumber\\
&\times&\frac{{}_2F_1(b_1^{\Tilde{k}}- b_3^{\Tilde{k}}+2,b_2^{\Tilde{k}}-b_3^{\Tilde{k}}+2;3-b_3^{\Tilde{k}};x)}    {{}_2F_1(b_1^{\Tilde{k}}- b_3^{\Tilde{k}}+1,b_2^{\Tilde{k}}-b_3^{\Tilde{k}}+1;2-b_3^{\Tilde{k}};x)} . \label{eq::h_k_NO_vac}
\end{eqnarray}
The mode solution is then determined using Eq. \eqref{muh}. In this way, we select the vacuum solution in the P\"oschl-Teller epoch, continuing it until the end of inflation by the procedure explained in the preceding section. With its value there, we can calculate the PPS.\\

\section{Primordial power spectrum}

As we have commented, the PPS for tensor perturbations is given by \cite{Langlois1}
\begin{eqnarray}
\mathcal{P}_\text{T} (k)= \frac{32k^3}{\pi } \frac{|u_k(\eta_{end})|^2}{a^2(\eta_{end})}.
\end{eqnarray}

We need just to introduce the solution found in de Sitter for the perturbation modes into this formula. On the other hand, compatibility with different types of observations requires that inflation must have produced at least around 60 e-folds \cite{Planck}, something that is fulfilled in the typical background solutions studied in LQC. Taking this fact into account, which ensures a large growth of the scale factor, it is reasonable to approximate the PPS as
\begin{eqnarray}
\mathcal{P}_\text{T} (k)= \frac{32k^3}{\pi} \lim_{a\to\infty} \left(\frac{|u_k|^2}{a^2}\right)
= \frac{16 H_\Lambda^2}{\pi} \left|F_k -G_k\right|^2.
\end{eqnarray}

The PPS critically depends on the integration constants $F_k$ and $G_k$, that fix the mode solutions for the perturbations. So, the behavior of the spectrum depends on this choice of solution. The form of the power spectrum in the above expression generally leads to fast oscillations in $k$ because, even if the norms of $F_k$ and $G_k$ vary slowly, their phases $\theta_k^F$ and $\theta_k^G$ generically produce a rapidly oscillating interference. Concretely,  
\begin{equation}\label{PPSosci}
\mathcal{P}_\text{T}(k) = \frac{16 H_\Lambda^2}{\pi} \Big[|F_k|^2 + |G_k|^2 - 2 |F_k||G_k| \cos{(\theta_k^F - \theta_k^G)}\Big].
\end{equation}
Actually, even if we start with initial conditions at the bounce for a non-oscillating vacuum state (of NO-AHD type), rapidly changing phases in $F_k$ and $G_k$ can be generated by the instantaneous transitions between different epochs in our matching of solutions, which are just a simplifying approximation where the smoothness of the effective mass is lost \cite{NM} (recall that, in the best of cases, we have preserved at most the continuity up to the first derivative in these matches). However, it is possible to remove such spurious oscillations in $k$, which on average would artificially pump power to the spectrum. Indeed, as it was argued and supported in Ref. \cite{NM}, we can always modify our solution as follows:
\begin{eqnarray}
F_k \to \Tilde{F}_k = |F_k|, \quad \quad G_k \to \Tilde{G}_k = |G_k|.
\end{eqnarray}
The above change of solution is in fact a Bogoliubov transformation \cite{NM}. The corresponding PPS is then
\begin{eqnarray}
\mathcal{P}_\text{T}(k) = \frac{16 H_\Lambda^2}{\pi} \left(|G_k| - |F_k|\right)^2 .\label{eq::PW_formula_cstes}
\end{eqnarray}
Notice that this spectrum is the envelope of the minima of the oscillating one in Eq. \eqref{PPSosci}. This redefined PPS is clearly free of the commented rapid oscillations but keeps all the main information of the original spectrum. 

As we have seen, to compute the PPS for our cosmological model we need initial conditions for the perturbation modes, or equivalently, to fix a vacuum state. Our choice is the NO-AHD vacuum, because it is optimally adapted to the background dynamics around the bounce and avoids spurious oscillations in that epoch. As for the corresponding initial conditions, provided e.g. at the conformal time $\eta_0$, from Eq. \eqref{muh} we can deduce \cite{NM}
\begin{eqnarray} \label{eq::u_k_NO_vac}
u_k(\eta_0) = \sqrt{-\frac{1}{2 \text{Im}(h_k)(\eta_0)}}, \quad u'_k (\eta_0)= - h_k^*(\eta_0)u_k(\eta_0). \label{eq::u_k_NO_vac}
\end{eqnarray}
Therefore, we only need the expression of $h_k$ picked out by the NO-AHD criterion (evaluated at $\eta_0$). We can then use our matching formulas to obtain the value of the constants $F_k$ and $G_k$ in the inflationary epoch. These constants are given by Eqs. \eqref{eq::F_k_deSitter} and \eqref{eq::G_k_deSitter}. With them, finally, we can calculate the PPS using Eq. \eqref{eq::PW_formula_cstes}. 

It is then straightforward to see, in particular, that in the ultraviolet regime ($k\gg1$) the asymptotic behavior of the hypergeometric function $_2F_1(a_1,a_2;a_3;x)$ for large $a_3$ \cite{Abra} implies that
\begin{eqnarray}
\text{Im}(h_k) = - k + \mathcal{O}(k^{-1}), \quad \text{Re}(h_k) = \mathcal{O}(k^{-2}),
\end{eqnarray}
and therefore
\begin{eqnarray}
u_k(\eta_0) = \sqrt{\frac{1}{2k}} + \mathcal{O}(k^{-5/2}), \quad u_k'(\eta_0)= -i \sqrt{\frac{k}{2}}+\mathcal{O}(k^{-3/2}).
\end{eqnarray}
Then, following our matching procedure, we find in the de Sitter epoch that
\begin{eqnarray}
|G_k|\approx\mathcal{O}(k^{-2}) \quad \text{and} \quad |F_k|\approx 1+\mathcal{O}(k^{-2}).
\end{eqnarray}
Hence, the PPS is scale invariant in the ultraviolet region, as in the standard cosmological model for the Bunch-Davies vacuum (in consonance with the CMB observations, extrapolating the discussion from tensor to scalar perturbations by assuming that the inflaton potential does not play a crucial role for these issues).

\begin{figure}[h!]
\centering
\includegraphics[width=16cm]{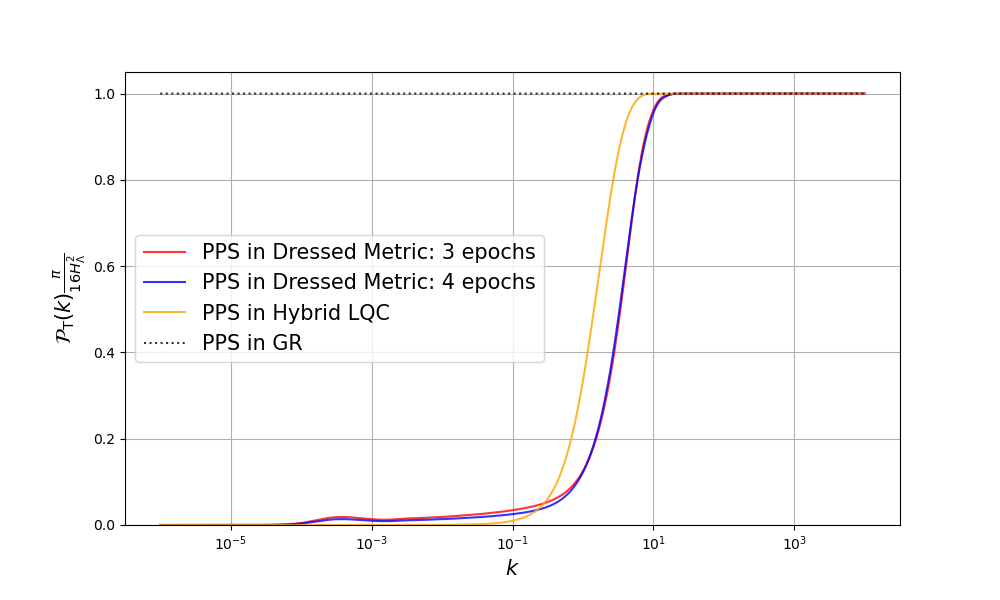}
\caption{PPS of the NO-AHD vacuum in the hybrid prescription for LQC (orange), in the dressed metric prescription for LQC including a constant mass epoch in the approximation (blue), and in the same prescription for LQC but without this last epoch (red). We also display the PPS for the Bunch-Davies state in GR (dark). } 
\label{fig::PW_log}
\end{figure}
\begin{figure}[h!]
\centering
\includegraphics[width=16cm]{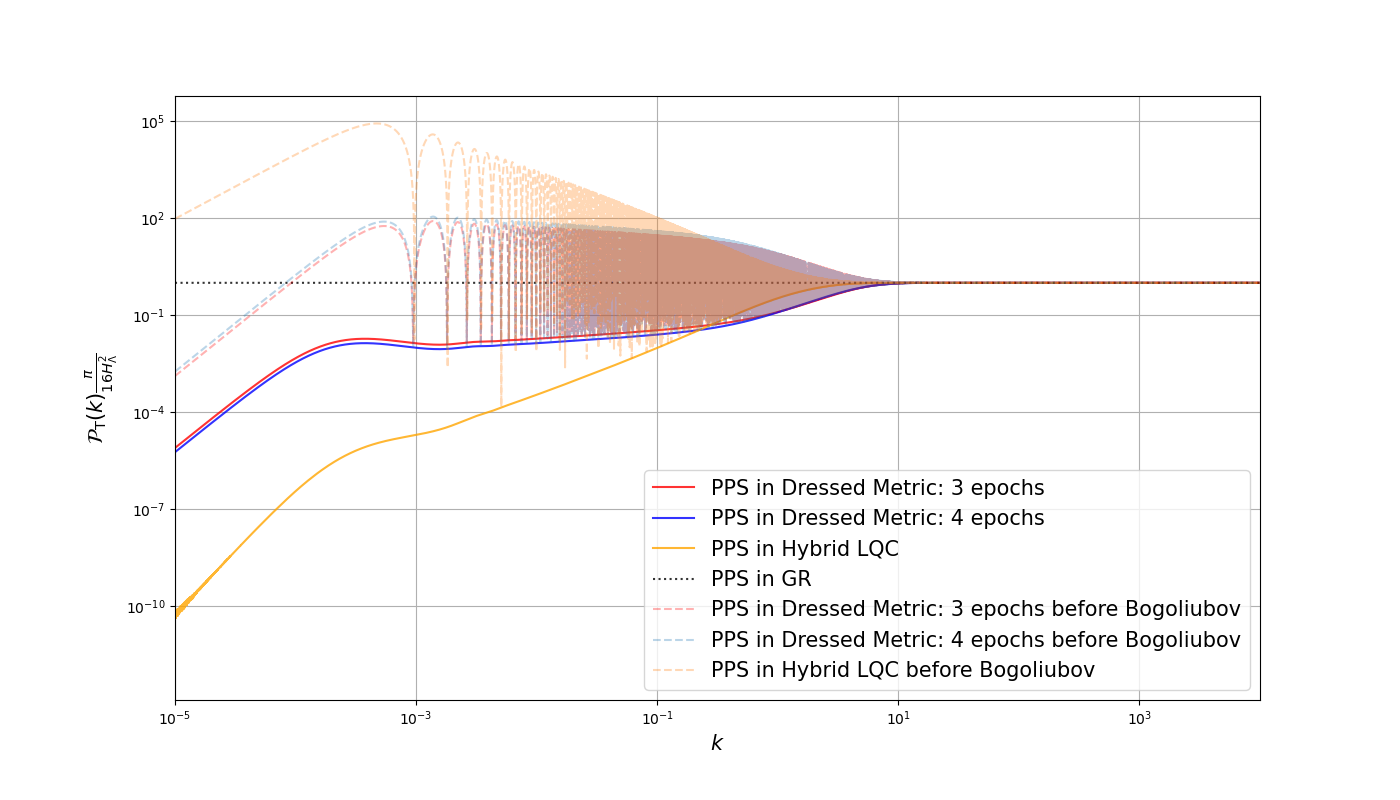}
\caption{PPS using the same color code as in the left figure. Dashed lines correspond to the PPS before applying the Bogoliubov transformation that removes spurious oscillations, whereas solid lines are used for the PPS when the Bogoliubov transformation is applied. }
\label{fig::PW_log_log}
\end{figure}

Moreover, employing the expression of $h_k$ for all $k$ given in Eq. \eqref{eq::h_k_NO_vac}, and using Eq. \eqref{eq::u_k_NO_vac} and our matching formulas, we can compute numerically the PPS and compare it, on the one hand, with the power spectrum expected in GR for a Bunch-Davies vacuum and, on the other hand, with the PPS for the same vacuum proposal in LQC adopting the hybrid prescription \cite{NM} to determine the background-dependent mass of the perturbations \cite{NBMmass}. In the following, we particularize the PPS of our model to the case with the usual value of the parameter $D$ in LQC, so that this PPS can be directly considered as the prediction for the dressed metric prescription with the standard value of the critical density. To check the robustness of our approximations, we include another PPS for this dressed metric, namely, the spectrum obtained by removing the constant effective mass epoch from our considerations and reducing to three the number of epochs considered in the evolution. In this alternative approximation, we still demand continuity up to the first derivative of the background and of the mode solutions, but we renounce to a continuous effective mass at the matching between the P\"oschl-Teller and the kinetically dominated epochs. 

These PPS are shown in Figs. \ref{fig::PW_log} and \ref{fig::PW_log_log}. We see in Fig. \ref{fig::PW_log} that the PPS of the NO-AHD vacuum has in practice a cutoff in the dressed metric prescription for LQC, case which also applies to the modified cosmology derived from thermodynamic considerations in the sense explained above, while the PPS of a Bunch-Davies state in GR is fully scale invariant. We observe that the cutoff wavenumber scale for the dressed metric prescription is of the same order but slightly greater than the corresponding one for hybrid LQC \cite{NM}. This provides some more power suppression for wavenumbers approximately greater than $0.2$. However, we note that the suppression is steeper in hybrid LQC for wavenumbers below its respective cutoff. Moreover, further in the infrared region, the PPS for a NO-AHD vacuum in hybrid LQC displays a considerably greater power suppression than the other two scenarios. On the other hand, in the dressed metric prescription for LQC, Figs. \ref{fig::PW_log} and \ref{fig::PW_log_log} show that there is just a very little bit more of suppression when the evolution is approximated using four different regimes than when these are only three, neglecting the constant mass period, but the difference is really tiny. The cutoff scale, however, is approximately the same in both cases.

\begin{figure}[h!]
\centering
\includegraphics[width=16cm]{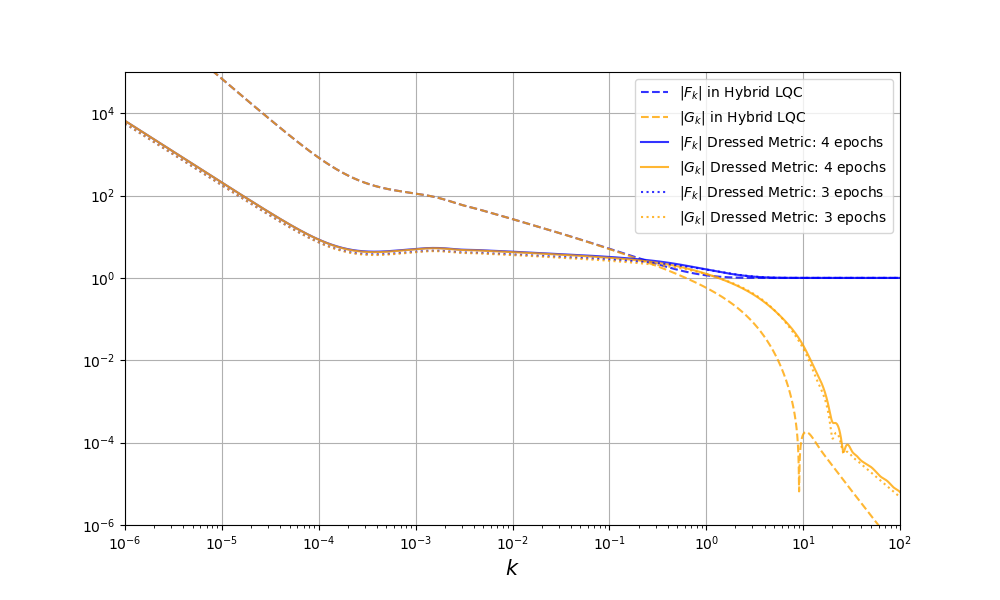}
\caption{Norms of the constants $F_k$ and $G_k$ for the different cases considered in LQC. The dashed, solid, and dotted lines correspond respectively to the hybrid prescription, the dressed metric prescription including a constant mass epoch, and the dressed metric prescription without this epoch. }
\label{fig::PW_F_k}
\end{figure}
\begin{figure}[h!]
\centering
\includegraphics[width=16cm]{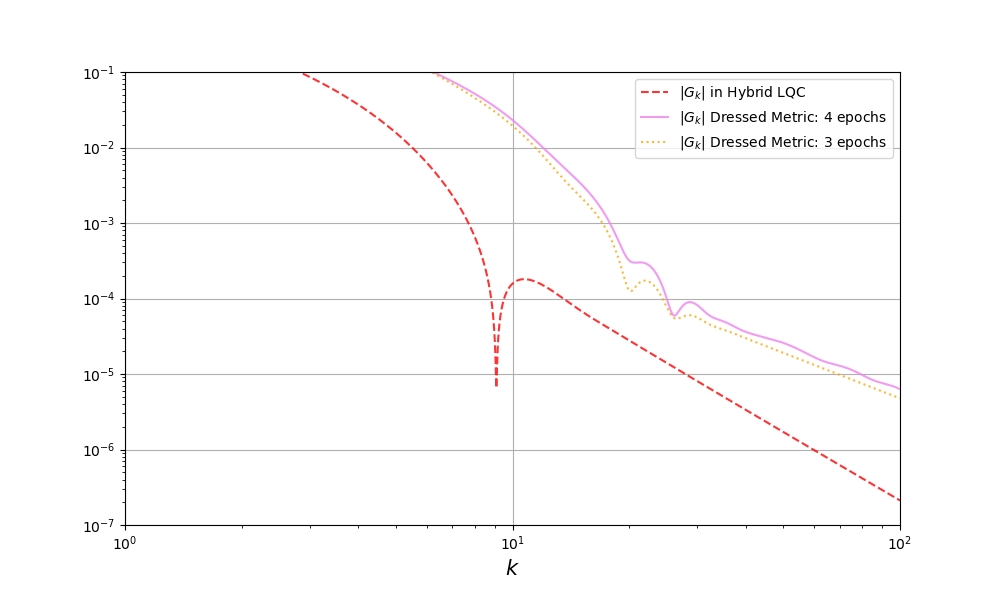}
\caption{Zoom-in of Fig. \ref{fig::PW_F_k}, showing the region around $k\sim10$. }
\label{fig::PW_G_k_zoom}
\end{figure}

\begin{figure}[h!]
\centering
\includegraphics[width=16cm]{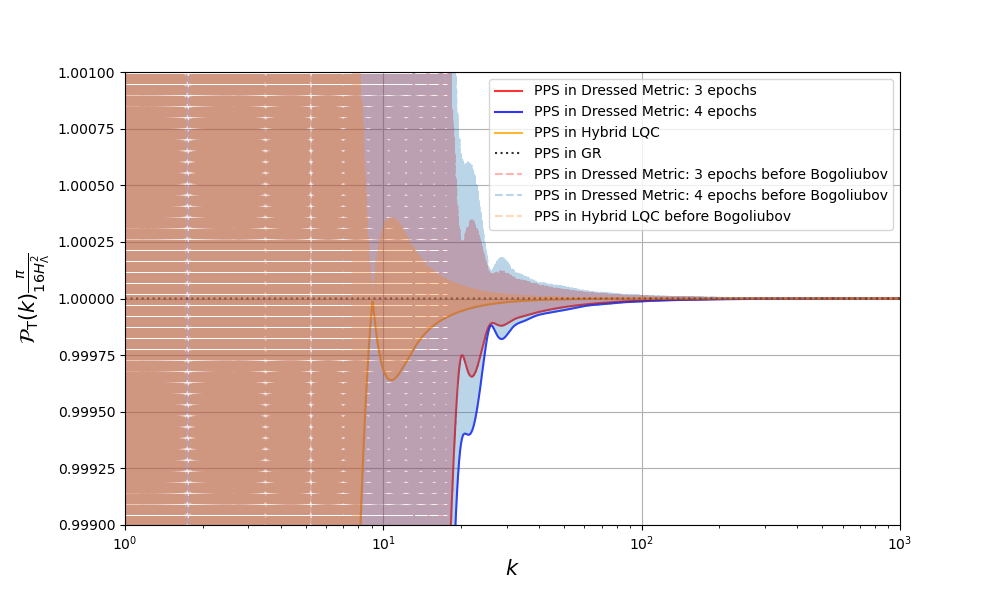}
\caption{Zoom-in around $k\sim 10$ of the PPS  in the hybrid prescription for LQC (orange), the dressed metric prescription for LQC with a constant mass epoch (blue), the dressed metric prescription for LQC without this epoch (red), and a Bunch-Davies state in GR (dark).}
\label{fig::PW_log_log_zoom}
\end{figure}
\begin{figure}[h!]
\centering
\includegraphics[width=16cm]{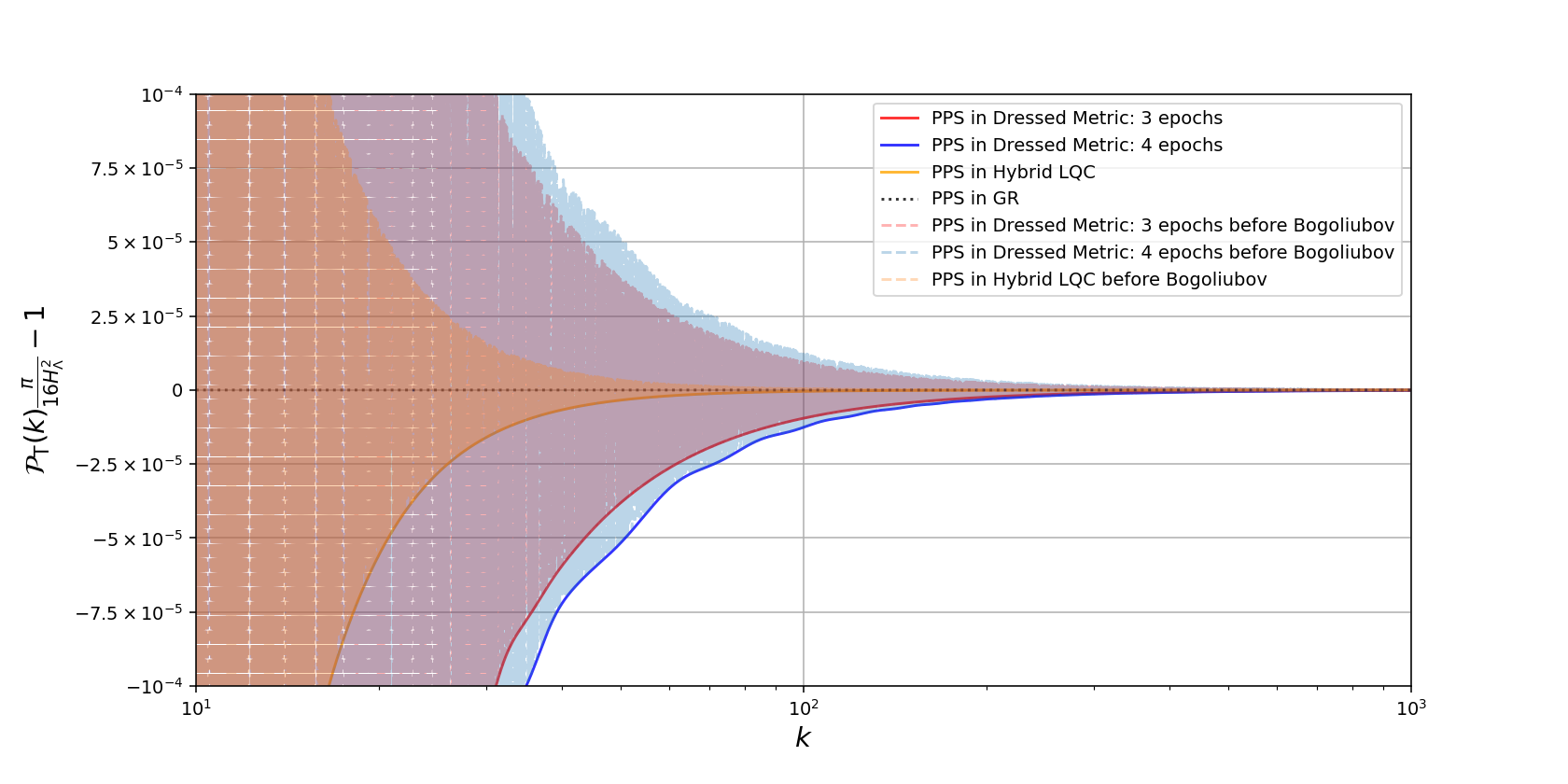}
\caption{Zoom-in around $k\sim 100$ of the PPS, showing the difference $[\mathcal{P}_\text{T}\pi^2/(16 H_{\Lambda}^2)]-1$, in the same cases and with the same color code as in Fig. \ref{fig::PW_log_log_zoom}. Units in the vertical axis are expressed in multiples of $10^{-5}$.}
\label{fig::PW_log_log_zoom2}
\end{figure}

The behavior of the constants $F_k$ and $G_k$ in Figs. \ref{fig::PW_F_k} and \ref{fig::PW_G_k_zoom} confirms our previous prediction that $F_k \approx 1$ in the ultraviolet region. Additionally, the constant $G_k$ becomes negligible in that region. However, it displays distinct asymptotics in the two considered prescriptions for LQC. In contrast, in the dressed metric prescription, the asymptotic behavior does not change significantly with the approximation adopted for the effective mass, with either three or four epochs, in spite of the fact that the number of matches and the continuity properties at them depends on this approximation. In Figs. \ref{fig::PW_log_log_zoom} and \ref{fig::PW_log_log_zoom2}, we display in more detail the region close to the cutoff in the PPS. The clear similarities between the results of the two considered approximations for the dressed metric prescription prove the robustness of our analysis.

\section{Conclusion}

We have investigated the cosmological implications of the dressed metric prescription for LQC. More specifically, we have discussed the influence of the quantum geometry effects on the PPS of cosmological perturbations, especially of tensor ones. We have started with a description of the effective cosmological backgrounds of interest in LQC. These backgrounds display a bounce at curvature scales of Planck order. In addition, for backgrounds of phenomenological interest, for which the quantum modifications are expected to be compatible with observations but still not physically irrelevant \cite{Morris}, we can ignore most of the effects of the inflaton potential on the evolution of the perturbations, assuming first dynamics where the kinetic contribution of the inflaton dominates, with negligible potential, and afterward an inflationary period of de Sitter expansion, as it would happen if the inflaton potential remained constant during this stage. In these circumstances, the perturbations evolve as if they were tensor ones. In this sense, our study can be interpreted to be focused on the PPS of tensor perturbations. Moreover, according to the dressed metric prescription, the perturbations evolve similarly as in GR but incorporating the quantum geometry modifications in the background evolution. The name {\it{dressed metric}} precisely reflects this procedure, because the background metric is dressed with quantum corrections, in our treatment using the effective description of LQC. 

We have also noted that the same type of background cosmology is found in a phenomenological theory of gravity with quantum modifications, rooted in thermodynamics, with the only difference that in this case the critical density at the bounce can be viewed as a free phenomenological parameter, not necessarily of Planck order as in LQC. Details about this coincidence are discussed in the Appendix, and involve a truncation of the thermodynamic equations for cosmology at first subleading order in a perturbative approach (for another approach developed in the same thermodynamic framework, see Ref. \cite{Cesare}). As a consequence, at least for values of the critical density close to its LQC value, our treatment of the perturbations extends to this model based on thermodynamic corrections if we assume that the only relevant effect of those corrections in the propagation of the perturbations is via the modification of the background, as it is advocated by the dressed metric proposal. 

In the above setting, we have obtained the exact form of the background-dependent effective mass that affects the propagation of the perturbations. Nonetheless, we have found obstructions to express it analytically in terms of the conformal time, which is the natural time that arises in the propagation equations of the perturbation modes. To circumvent these obstructions and derive also analytically the form of the general solution to these equations, we have employed certain approximations in which the evolution is divided into several epochs, with different dominating regimes in each of them. We have considered an approximation with four epochs (a bounce with a P\"oschl-Teller mass, an interval with constant effective mass, a kinetically dominated period, and a de Sitter stage), as well as an alternative approximation with only three of those epochs (eliminating the interval with constant effective mass). Comparison between these two approximations has allowed us to put to the test the robustness of the results. The matching between different epochs in our approximations has been performed by demanding that both the background and the perturbation modes are continuous up to their first derivatives in the transition.

Since the considered backgrounds are nonstationary and their isometries generically do not involve the time direction, in principle there is no preferred state that can serve as a natural vacuum to construct a quantum field theory for the perturbations. In this situation, a physical criterion is needed to select a vacuum, other than the invariance under symmetries. With this aim, we have adhered to the NO-AHD proposal. This proposal picks out a state with good non-oscillating properties that can be specified by an asymptotic diagonalization in the ultraviolet of the Hamiltonian that governs the (unitary Heisenberg) dynamics of the perturbations, typically in the epoch with relevant quantum effects \cite{NMT,NM}. The non-oscillating behavior shows that the state is optimally adapted to the background dynamics, and translates into similar properties in the corresponding PPS. Then, using the asymptotics in the ultraviolet that are characteristic of the NO-AHD proposal and extending the resulting $k$-dependence, we have determined the initial conditions on the perturbation modes that correspond to our choice of vacuum state.

With all these ingredients, finally, we have computed the PPS of the NO-AHD vacuum state in the dressed metric prescription, with the standard value of the critical density in LQC. We have compared this PPS with the spectrum obtained for the same kind of vacuum in LQC with the alternative, hybrid prescription for the quantization of the perturbations, as well as with the scale invariant spectrum obtained in standard GR with a Bunch-Davies vacuum in de Sitter spacetime. 

The main feature of the obtained PPS is the appearance of power suppression with respect to the standard GR case at an effective cutoff that is of Planck order (however, it is worth noting that the corresponding physical scale can currently be observable in the CMB \cite{NM}). Our analysis shows a small variation of this cutoff scale when the model is compared with a hybrid model in LQC (both with the same energy density at the bounce), adopting a NO-AHD vacuum for the perturbations. For wavenumbers greater than (approximately) $0.2$, this leads to some more power suppression in the case of the dressed metric model, but remarkably the suppression is steeper in the hybrid approach for this kind of vacuum state. In particular, further in the infrared it becomes greater than for the dressed metric model. This power suppression may depend on the specific approximation adopted to describe the background-dependent mass of the perturbations, but we have seen that the change is small. In this sense, our results are robust, obtaining essentially the same conclusions for an approximation with four different dynamical epochs than for another one with only three of those epochs. In independent works, we plan to carry out a full numerical study of the PPS, integrating the mode solutions without analytic techniques based on approximations to their effective mass. This will shed valuable light on the validity of those approximations. On the other hand, our study can also be extended to other positive values of the critical density (or, equivalently, of $D$) almost straightforwardly, because the methods developed in this work are mostly parameter-independent. In this way, one can investigate the properties of the PPS for other, more general cosmologies, corresponding to the phenomenological model rooted on thermodynamic considerations. It would be very interesting to perform a more complete analysis constraining the observationally admissible values of $D$. Another appealing possibility for further research would be to compare the power spectrum calculated by these means for the NO-AHD vacuum with the spectra corresponding to other vacua, e.g. adiabatic states \cite{Parker,Lueders}, the vacuum suggested by Ashtekar and Gupt \cite{AG1,AG2} (motivated by minimal quantum uncertainties near the bound and some classical behavior after inflation), or the low-energy state proposed in Ref. \cite{Benito}.    

\section*{Acknowledgments}
This work was partially supported by Project No. MICINN PID2020-118159GB-C41 from Spain. AA-S is funded by the Deutsche Forschungsgemeinschaft (DFG, German Research Foundation) -- Project ID 516730869. AV-B acknowledges CSIC by financial aid under the scholarship JAEINT\_22\_01489. The authors are greatly thankful to Beatriz Elizaga Navascu\'es, who made the first analyses of the behavior of the power spectra for the models and approximations studied in this work. She shared many enlightening ideas and maintained useful discussions with the authors. They are also grateful to Jes\'us Y\'ebana for his help with the numerics and a fruitful interchange of ideas, to Marek Li\v{s}ka for all the discussions along the development of the project, and to Mercedes Mart\'{\i}n-Benito, Rita Neves, and Javier Olmedo for conversations. 

\appendix

\section{Connection with Termodynamics of Spacetime} 

The thermodynamics of spacetime has been attracting an increasing interest as an alternative approach to understand the ultimate description of gravity. Among the reasons that explain this interest, we can mention the development of black hole thermodynamics \cite{Bekenstein} and the recent derivation of the Einstein equations from thermodynamical properties of the spacetime \cite{Jacobson1,Jacobson2}. These advances have contributed to the innovative idea of encoding the gravitational dynamics into thermodynamics \cite{Jacobson2,Alonso1}. 

By modifying the thermodynamics of spacetime to include low-energy quantum gravity effects, phenomenological equations have been put forward to correct the gravitational dynamics, recovering the Einstein equations in the classical limit \cite{Alonso2}.  These modified gravitational equations have been derived from a condition of entanglement equilibrium imposed on local causal diamonds, constructed at every spacetime point. Since the variation of the total entropy defining the equilibrium condition is the sum of the variations of the matter entanglement entropy, $\delta S_{\text{m}}$, and of the entanglement entropy of the horizon, $\delta S_{\text{B}}$ (which corresponds to the modified Bekenstein entropy \cite{Jacobson2,Alonso2}), the condition becomes $\delta S_{\text{B}} + \delta S_{\text{m}} =0$. To study (low-energy) quantum modifications to the gravitational dynamics, one considers the leading order correction to the Bekenstein entropy, given by a term that is logarithmic in the horizon area $\mathcal{A}$,
\begin{equation}\label{eq::modi-entrop}
S_{\text{B}}=\frac{\mathcal{A}}{4}+30\pi D_{{\rm th}} \ln{\left(\frac{\mathcal{A}}{\mathcal{A}_0}\right)}+ O\left(\frac{1}{\mathcal{A}}\right).
\end{equation}
Here, $\mathcal{A}_0$ is an arbitrary constant with dimensions of area and $D_{{\rm th}}$ is a real dimensionless parameter that depends on the specific quantum gravity theory employed to calculate the modifications to the Bekenstein entropy. We will treat it as a free parameter in the rest of our discussion. 

The above logarithmic form of the entropy correction is nearly universal inasmuch as it is predicted by a variety of candidate theories of quantum gravity (e.g. string theory \cite{Sen}, LQG \cite{Kaul}, or AdS/CFT correspondence \cite{Faulkner}), as well as by many model-independent phenomenological arguments \cite{Adler,GM,Gour,Davidson} and entanglement entropy calculations \cite{Mann}.

The matter entanglement entropy crossing the horizon for small perturbations can be expressed in terms of a local modular Hamiltonian proportional to the stress-energy tensor (see Ref. \cite{Jacobson2} for details). Using the expressions of both entropy variations, one can then find the traceless modified equations for the gravitational dynamics \cite{Alonso2}. One can apply these general equations to the case of an FLRW cosmology with a (free) scalar field to obtain the modified Raychaudhuri equation
\begin{eqnarray}\label{eq::Raychaudhuri}
\Dot{H}-\frac{\kappa}{a^2} - D_{{\rm th}} \left(\Dot{H} - \frac{\kappa}{a^2} \right)^2 = -4\pi  \dot{\phi}^2 \label{eq::Mod_Ray_with_curv},
\end{eqnarray}
where $\kappa$ is the spatial curvature. Compared to the standard equation for FLRW in GR, one recognizes an extra term proportional to the correction parameter $D_{{\rm th}}$. Therefore, the limit $D_{{\rm th}} \to 0$ reproduces the Raychaudhuri equation in the classical model \cite{Alonso3}. Assuming local energy conservation, restricting the attention to the spatially flat case for simplicity, and using perturbative methods (with perturbative corrections truncated at first subleading order), one reaches the modified Friedmann equation \cite{Alonso3}
\begin{eqnarray}\label{eq::Modified_Friedmann_free_field}
H^2 = \frac{4\pi  }{3} \dot{\phi}^2  \Big [1 - 2\pi D_{{\rm th}}  \dot{\phi}^2 \Big] .
\end{eqnarray}
This equation is entirely similar to that found in effective LQC. In fact, the thermodynamics of the spacetime developed in Ref. \cite{Alonso3} recovers the same cosmological solutions as in effective LQC if one identifies the phenomenological parameter $D_{{\rm th}}$ with the LQC parameter $D$, and hence sets it equal to $1/(4\pi\rho_c)$.

\end{document}